\documentclass{aa}
\usepackage{graphicx}
\usepackage{threeparttable}
\newcommand{\ha}{H$\alpha$}
\newcommand{\hb}{H$\beta$}

\usepackage{txfonts}
\usepackage{hyperref}
\usepackage{adjustbox}

\hypersetup{
    colorlinks=true,
    linkcolor=blue,
    filecolor=magenta,      
    urlcolor=cyan,
    citecolor=blue,
}
\begin{document} 

\title{X-ray view of emission lines in optical spectra: Spectral analysis of the two low-mass X-ray binary systems Swift J1357.2-0933 and MAXI J1305-704}
   
\author{A. Anitra\inst{1}, C. Miceli\inst{1,2,3}, T. Di Salvo\inst{1}, R. Iaria\inst{1}, N. Degenaar\inst{6}, Jon M. Miller \inst{7}, F. Barra \inst{1}, W. Leone \inst{1,4}, L. Burderi\inst{5}}

   \institute{Università degli Studi di Palermo, Dipartimento di Fisica e Chimica, via Archirafi 36, I-90123 Palermo, Italy
   \and 
   INAF/IASF Palermo, via Ugo La Malfa 153, I-90146 Palermo, Italy
   \and
   IRAP, Universitè de Toulouse, CNRS, UPS, CNES, 9, avenue du Colonel Roche BP 44346 F-31028 Toulouse, Cedex 4,France
   \and
     Department of Physics, University of Trento, Via Sommarive 14, 38122 Povo (TN), Italy 
    \and
    Dipartimento di Fisica, Universit\`a degli Studi di Cagliari, SP Monserrato-Sestu, KM 0.7, Monserrato, 09042 Italy 
    \and
    Anton Pannekoek Institute for Astronomy, University of Amsterdam, Postbus 94249, 1090 GE Amsterdam, The Netherlands 
    \and
    Department of Astronomy, The University of Michigan, 1085 South University Avenue, Ann Arbor, MI, 48109, USA}

 
\abstract{
We propose a novel approach for determining the orbital inclination of 
low-mass X-ray binary systems by modelling the \ha{} and \hb{} line profiles emitted by the accretion disc, with a Newtonian (i.e. non-relativistic) version of \textsc{diskline}.
We applied the model to two sample sources, Swift J1357.2-0933 and MAXI J1305-704, which are both transient black hole systems, and analyse two observations that were collected during a quiescent state and one observation of an outburst. The line profile is well described by the \textsc{diskline} model, although
we had to add a Gaussian line to describe the deep inner core of the double-peaked profile, which the \textsc{diskline} model was unable to reproduce. 
The \hb{} emission lines in the spectrum of Swift J1357.2-0933 and the \ha{} emission lines in that of MAXI J1305-704 during the quiescent state are consistent with a scenario in which these lines originate from a disc ring between $(9.6-57) \times 10^{3}, \rm{R_{g}}$ and $(1.94-20) \times 10^{4}, \rm{R_{g}}$, respectively.

We estimate an inclination angle of $81 \pm 5$ degrees for Swift J1357.2-0933 and an angle of $73 \pm 4$ degrees for MAXI J1305-704. This is entirely consistent with the values reported in the literature. 

\indent In agreement with the recent literature, our analysis of the outburst spectrum of MAXI J1305-704 revealed that the radius of the emission region deviates from expected values. It is larger than the orbital separation of the system.
 This outcome implies several potential scenarios, including line profile contamination, an alternative disc configuration that deviates from the Keplerian model, or even the possibility of a circumbinary disc.
\\
\indent We caution that these results were derived from a simplistic model that may not fully describe the complicated physics of accretion discs. 
Despite these limitations, our results for the inclination angles are remarkably consistent with recent complementary studies, and the proposed description of the emitting region remains entirely plausible.}

\titlerunning{An X-ray view of emission lines in optical spectra}
   \authorrunning{A.Anitra et al.}
   \maketitle
%
\section{Introduction}
Low-mass X-ray binaries (LMXBs) are systems that are composed of a compact object, such as a black hole (BH) or neutron star (NS), and a donor star with a mass lower than  1 $\rm{M_{\odot}}$. The compact object in these systems accretes matter from its companion via Roche-lobe overflow, which leads to the formation of an accretion disc around it.

The study of the double-peaked emission line profiles that are often present in the optical spectra of these sources is a powerful tool that allows us to determine the orbital parameters of the system. By examining parameters such as the orbital period and radial velocity semi-amplitude we can dynamically confirm the nature of the compact objects in these systems.

It has been acknowledged that the central core within the double-peaked lines that are emitted by an accretion disc become deeper with increasing inclination. This is particularly noticeable for inclinations greater than 67 degrees \citep{1986_Horne}.
\cite{Casares_2022} discovered a direct correlation between the depth of the inner core in the double-peaked H$\alpha$ emission line and the inclination angle in a group of quiescent BH LMXBs. By applying this method, the authors were able to determine the inclination angle of different sources with a high confidence level.

\cite{Wang_2009ApJ...703.2017W} examined the optical spectrum of the millisecond radio pulsar binary SDSS J102347.6+003841. They focused on the analysis of double-peaked emission lines to determine the accretion disc geometry. By adopting two Lorentz functions to fit the lines and calculating the flux, they were able to estimate key properties such as the disc temperature range, the outer and inner emission radii, and the overall mass of the disc.

In this work, we investigate whether an estimate of the system inclination angle can be derived by applying the \textsc{diskline} model for the first time to fit the double-peaked emission lines related to the H-series transition that appears in the optical spectra of two candidate transient BH systems. The \textsc{diskline} model can describe simple pure Keplerian fully symmetric disc profiles, where the emissivity is a pure power law. This may not be the case for optical lines that are emitted in the outer disc, which is most affected by asymmetries, non-uniform emissivity patterns, strong outer disc features such as hot spots and bulges, departures from Keplerian flow due to tidal interactions, and so on. However, we show that the line profiles observed during X-ray quiescence of the BH candidates Swift J1357.2-0933 and Maxi J1305-704 appear to be simple enough to provide useful constraints on their inclination angle.
\section{\label{section_2}X-ray transient system samples}
To ensure the accuracy of the inclination measurements, we chose two transient BH X-ray binaries that have been intensely studied in the literature and for which several indications of a high-inclination angle have been reported. 

\subsection{Swift J1357.2-0933}
Swift J1357.2-0933 (hereafter J1357) is an LMXB X-ray transient that was originally detected during its outburst in 2011. The distance to J1357 is estimated to be greater than 2.29 kpc, which places the source within the thick Galactic disc \citep{Daniel_2015}.
Studies of the modulation of the \ha{} double-peak emission line during 14 months of outbursts have led to an orbital period of $2.5673\pm 0.0006\, \rm{h}$ \citep{Casares_2022}. 
An estimate of the compact object mass was provided by \cite{Daniel_2015} based on the correlation between the full width at half maximum (FWHM) and the radial velocity of the donor star (K2) \citep{Casares_2015}. This revealed that this source is one of the most massive BHs in our galaxy ($\rm{M_{BH}}>9.3\, \rm{M_{\odot}}$). 
The optical light curve of the source displays periodic dips with a consistent pattern during the outburst \citep{Armas_2014}. These dips are thought to be linked to a toroidal structure within the inner accretion disc that gradually moves outward as the outburst proceeds \citep{Corral_santana_2013}.

The frequent dipping episodes in the system suggest a high-inclination angle $i$ larger than 70 degrees \citep[see][]{Corral_santana_2013,Torres_2015}.
\cite{Anitra2023} performed phase-resolved spectroscopy during a quiescent state. They focused on the analysis of the \hb{} line profile. The authors estimated the systemic velocity and the radial velocity semiamplitude of the black hole and detected compelling evidence of a narrow and variable inner core within the double-peaked \hb{} emission line profile.
These features were observed in high-inclination cataclysmic variables \citep{1983_Schoembs} and are probably associated with an occultation of the inner-disc emission by the outer rim bulge.

High-inclination systems typically exhibit eclipses in their light curves \citep{anitra_2021}. The absence of eclipses in this system appears to contradict a high inclination \citep[see][]{2018_iaria,Armas_2014}. \cite{Corral_santana_2013} justified this absence by considering an inclination angle of 78 degrees and a low mass ratio, proposing that the radius of the donor star might be comparable to or even smaller than the outer disc rim. Based on these indicators of a high-inclination nature, J1357 is an ideal candidate for our analysis. 

\subsection{MAXI J1305-704}
MAXI J1305-704 (hereafter J1305) was proposed to be a high-inclination (BH) X-ray binary because a dipping behaviour was detected during its outburst (\citealt{Suwa2012}, \citealt{Shidatsu_2013}, \citealt{Morihana2013}, \citealt{kennea2012}). The source has an estimated distance of $d=7.5_{-1.4}^l{+1.8} \, \rm kpc$, which places it in the thick Galactic disc \citep{2021_mata-sanchez}. 

J1305 was studied in detail in the X-rays, but follow-up at other wavelengths has been scarce \citep[see][]{Shaw2017,Miller_2014ApJ...788...53M,Shidatsu_2013}.
The first optical spectroscopic analysis of this source during its 2012 outburst has been performed by \citet{Miceli2023_submitted}. It focused on the double-peaked H$\alpha$ emission line, which revealed no conclusive evidence of the outflow features within the system.

\citet{2021_mata-sanchez} inferred an orbital period of  $9.456 \pm \, 0.096 \,\rm{h}$ and a $\rm{M_{BH}}=8.9^{+1.6}_{-1.0} \, \rm{M_{\odot}}$ by analysing the orbital modulation in the optical light curve in quiescence. The same authors obtained a constraint on the orbital inclination of $i=72_{-8}^{+5} \, \rm deg$, which supports the high-inclination scenario.
The number of observations during outburst and quiescence and the well-constrained values of the inclination angle and other orbital parameters (e.g. orbital period and mass of the two objects) make J1305 one of the best candidates for our analysis.

\section{Observations}
We analysed different sets of observations for the two sample sources.

J1357 was observed by the 10.4-meter Gran Telescopio Canarias (GTC) at the Observatorio del Roque de los Muchachos (ORM) on the island of La Palma (Spain), using the Optical System for Imaging and low-Intermediate-Resolution Integrated Spectroscopy (OSIRIS). 
Eight observations were collected during a quiescence state. They were focused on the wavelength range from 3950 to 5700 \AA. The R2000B grism was set alongside a 0.8" slit during the campaign. This configuration allowed us to obtain a spectral resolution of R = 1903 and a dispersion of D = 0.86 \AA/pix.
The dispersion is evaluated at $\lambda_{c} = 4755 $ \AA \,, and the resolution was obtained by the FWHM of the skylines in the background spectra.
Data were collected on 5 March 2016 between 03:32:11.4 and 06:19:5. Each observation was 1235 s long  for a total exposure of 2.75 h and covered a full orbit \citep{Anitra2023}.

J1305 was observed both during an outburst and in a quiescent state. 
The initial observation took place during its discovery outburst in 2012 and was made with the 6.5-meter \textit{Magellan} Clay telescope located at Las Campanas Observatory in Chile. This observation involved the use of the Low Dispersion Survey Spectrograph (LDSS-3) and the VPH-ALL grism setup.
The observations covered two consecutive nights, 2 May (02:18:34 UTC to 02:56:41 UTC) and 3 May (00:29:36 UTC to 01:09:03 UTC) with a 0.75'' slit. The exposure time for each individual spectrum was set at 300 s for a total of six spectra per night \citep{Miceli2023_submitted}. The setup configuration allowed us to focus on a wavelength range between 4500 - 6950 \AA  with a resolution power of R = 826. 

The source was observed during quiescence on 31 March 2016 with the Very Large Telescope Unit Telescope 1 (VLT-UT1; Paranal Observatory, Chile) using the focal reducer/low-dispersion spectrograph 2 \citep[FORS2,][]{1998_Appenzeller} in long-slit mode. The data sets were composed of 16 spectra, with an exposure time of 1800 s each. They were collected consecutively for a total exposure of $\sim \, 9 \, \rm{h}$, that is, they covered almost one orbital period \citep{2021_mata-sanchez}. 
The spectra cover a wavelength range between 5800 - 7300 \AA \, with a spectral resolution of R $\sim$ 2140 and a dispersion of 0.76 \AA/pix. This was obtained by measuring the FWHM of the skylines in the background spectra.

We reduced the data using standard procedures based on the \textsc{iraf\footnote{\textsc{iraf} is distributed by the National Optical Astronomy Observatories, operated by the Association of Universities for Research in Astronomy, Inc., under contract with the National Science Foundation.}} software, \textsc{molly} tasks, and \textsc{python} packages from \textsc{astropy} and \textsc{pyastronomy} \citep{astropy}. These tools allowed us to correct the observed spectra for bias and flats and to calibrate the data set. 
We note that cosmic rays contaminated the spectra of J1357, and we corrected for them using the L.A. Cosmic task \citep{2012_lacosmic}. We adopted the optimal extraction technique \citep{1998_Naylor} to extract the two-dimensional images.
The quiescence spectra of J1305 might be contaminated by the companion emission. However, as claimed by \cite{2021_mata-sanchez}, the contamination is probably lower than 10\% of the total flux and is therefore negligible.

\section{Analysis and results}
\subsection{Diskline model}\label{sec:disk}
The emission lines of the Balmer series are associated with the atomic hydrogen transition. They appear as broad symmetrical double-horned lines when observed from accretion discs at high-inclination angles. 
In these systems, the emission is mainly influenced by the Doppler shift, which is caused by the orbital motion of matter in the accretion disc around the compact object.

In a disc ring at a certain radius $\rm{R_{\mathrm{D}}}$, matter moves at the corresponding Keplerian velocity $\rm{V_{kep}}$.
For a distant observer, some of the matter within the disc moves toward the observer, while it simultaneously moves away from the observer on the far side of the disc. This dual motion causes the observer to perceive the emission line as simultaneously blue- and red-shifted, and it causes the distinctive symmetrical double-horn profile. Because the emission comes from different regions of the disc surface, the overall line profile is additionally broadened by the velocity distribution in the disc.
As first discussed by \cite{Smak_1981AcA....31..395S} and later by \cite{1986_Horne}, the final line profile is significantly affected by the inclination angle of the system relative to the line of sight because the component of the Keplerian velocity $\rm{V_{kep}}$ along the line of sight is $V_{\mathrm{D}}= \pm V_{\mathrm{Kep}}\left(R_{\mathrm{D}}\right) \sin i$. Furthermore, the separation of the peaks is determined by the outer radius of the emitting region, but the wings are shaped by the emissivity profile.


The X-ray spectrum-fitting software \textsc{xspec} \citep{xspec_1996} provides several models for describing an emission line that is shaped by Doppler effects from a Keplerian accretion disc. One of the most commonly adopted models for fitting double-horn emission lines is the \textsc{diskline} model \citep{fabian_1989}. It is usually adopted to describe the iron K$\alpha$ emission line that characterises the X-ray reflection features observed in LMXBs (see e.g. \citealt{DiSalvo_2009}).

The shape of the broad iron fluorescence line observed at X-ray wavelengths, similar to the hydrogen Balmer series lines in optical spectra, is affected by Newtonian Doppler shift. Its profile appears to be quite different, however.
The Fe K$\alpha$ line originates in the innermost part of the disc, where the velocity of matter reaches relativistic values. Consequently,  the relativistic beaming affects the emission and intensifies the blue peak of the line while weakening the red peak.  Furthermore, the line is gravitationally redshifted by the strong gravitational pull near the compact object, causing it to shift to lower energies in a way that depends on the distance from the compact object. As a result, the emission line appears to be broadened and asymmetric.

\begin{figure}
    \centering
    \includegraphics[width=.5\textwidth]{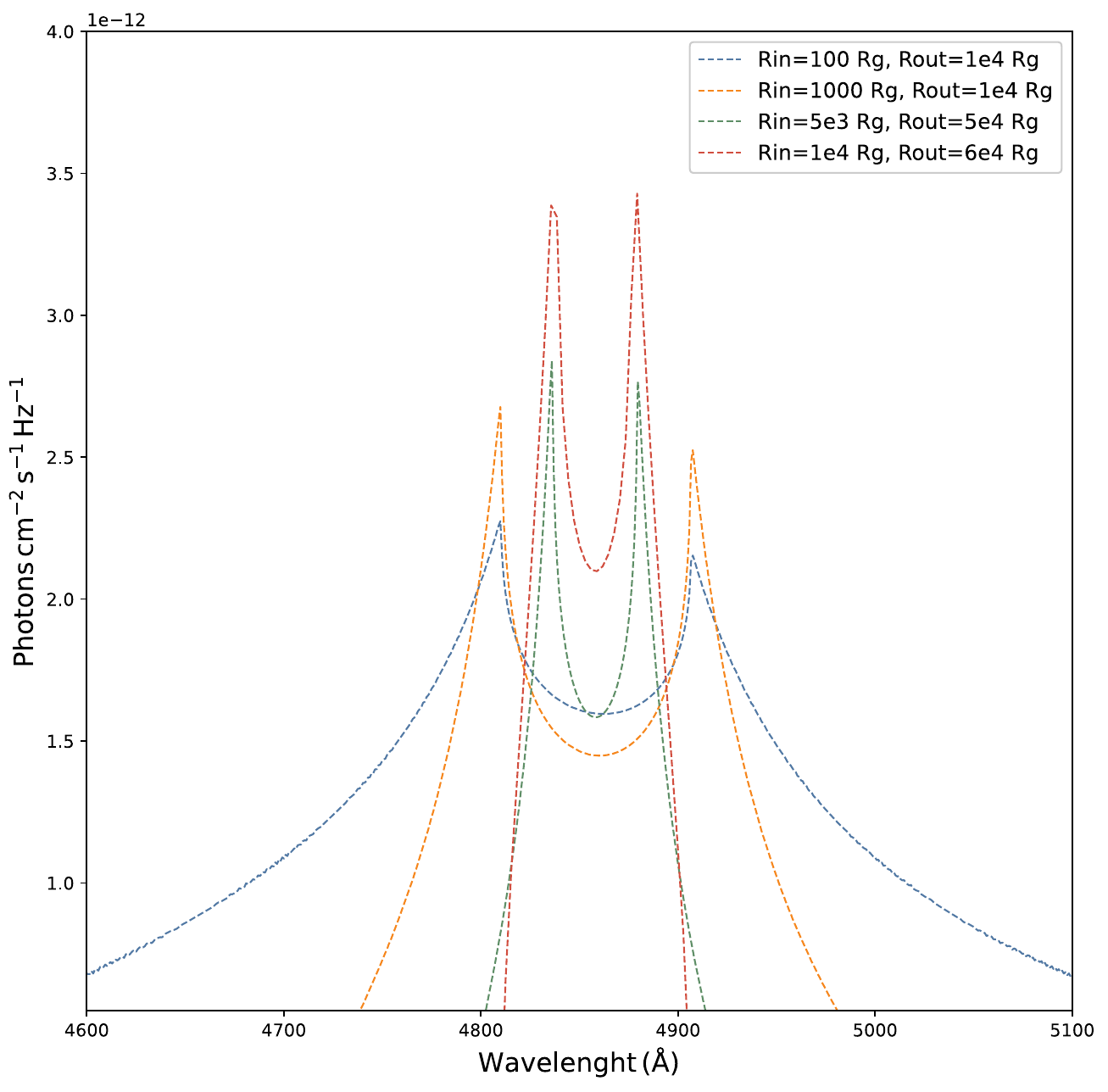}
    \caption{\label{shape_disk}Series of examples showing the \textsc{diskline} model profiles with fixed inclinations ($i$ = 83 degrees), emissivity index ($\beta$ = 2.5), and energy centroid (E = 4857.522 \AA) for all models. The inner and outer radius  were systematically varied within ranges of $(100 - 10^{4}) \,\rm{R_{g}}$ and $(10^{4} - 6\times10^{4}) \,\rm{R_{g}}$, respectively, to highlight the variations in the line profiles. The model normalisations are scaled for visual clarity. }
\end{figure}

The \textsc{diskline} model is composed of the following free parameters that shape the emission line profile: the inner ($\rm{R_{in}}$) and outer radius ($\rm{R_{out}}$) of the emission region in the disc (expressed in units of gravitational radii, $\rm{R_{g}=GM/c^{2}}$), the orbital inclination of the system, and the index, $\beta$, which describes the dependence of the disc emissivity on the distance from the compact object (E $\propto R^{\beta}$). This model was used several times to obtain constraints on the inclination angle of LMXBs \citep[see e.g.][]{anitra_2021,2009_iaria,Cackett_2008} when relativistic effects were dominant.
Although this model is usually adopted to describe a (relativistic) emission line, it can be adapted to model the hydrogen line profiles as well. When the value of the inner and outer radius is high enough, meaning that the emission region of the line is far from the compact object, the relativistic beaming and the gravitational redshift indeed become negligible, and only the Newtonian Doppler shift dominates (see Fig. \ref{shape_disk}). 
Under these conditions, the model is equivalent to the optically thin solution presented in \citet{1986_Horne}. 
The outer radius of the disc is also most strongly affected by asymmetries, non-uniform emission line emissivity patterns, and strong outer disc features. For instance, self-absorption by an optically thick disc atmosphere can contribute to deepening the inner core of the line. Simultaneously, disc precession may play a significant role in determining the orbit-averaged velocity shifts that are observed in the emission line centroid \citep{Torres_2002}, while a hot spot can enhance one of the two peaks of the line \citep{Marsh_Robinson_1994MNRAS.266..137M}. 
For these reasons, it is important to exercise caution with respect to the constraints derived from the analysis.

In order to isolate the emission line profile on which we focused our analysis, we normalised the spectra of J1357 and J1305 by dividing them by their polynomial best-fit functions.
Specifically, we applied a third-order polynomial fit for the first source and a fifth-order polynomial fit for the second source. We thus adapted our optical data sets for an analysis in the \textsc{xspec} environment by creating a unit diagonal response matrix of the data size. 
We adhered to the following naming convention: spectra linked to J1357 are denoted src1-Q, and those associated with J1305 are labelled src2-O during outburst and src2-Q during quiescence.

We find it convenient to express the constrained best-fit radii in units of the expected tidal radius of the system, that is, the outer edge of the disc truncated by the tidal torque of the companion star \citep{King}. The tidal radius depends on several binary parameters, but its value can be shown to be close to $\rm{R_{T}}=0.9\, \rm{R_{1}}$ \citep{King}, where R$_{1}$ is the Roche-lobe radius of the compact object. This can be calculated using the \cite{Eggleton_1983} equation
\begin{equation}\label{roche}
    \frac{R_{1}}{a}=\frac{0.49q^{-2/3}}{0.6q^{-2/3}+\ln{(1+q^{-1/3})}},
\end{equation}
where q is the mass ratio of the two bodies $\rm{M_{1}}$ and $\rm{M_{2}}$, and a is the orbital separation of the system, which can be calculated using Kepler's third law. Moreover, adopting black hole masses of about 8.9 solar masses for J1305 \citep{2021_mata-sanchez} and 9.3 solar masses for J1357 \citep{Daniel_2015}, we can define the  gravitational radius $R_{g}$ for the two systems as 13 km and 14 km, respectively.
Consequently, we obtain an orbital separation and tidal radius  of  $\rm{a}=(1.0259 \pm 0.0002)\times 10^{6} \, \rm{R_{g}}$  and $\rm{R_{T}} = (5.9\pm 0.004) \times 10^{4} \, \rm{R_{g}}$ for J1357,  and  $\rm{a}=(2.524 \pm 0.002)\times 10^{5}\, \rm{R_{g}}$, and $\rm{R_{T}}  = (1.45\pm 0.06) \times 10^{5} \, \rm{R_{g}}$ for J1305.

\subsection{J1357 spectroscopy}
The wavelength coverage of our spectra allowed us to focus on the H$\beta$ line region with a sufficiently high signal-to-noise ratio (S/N)\footnote{The H$\gamma$ line was also detected, but its lower S/N combined with a low line intensity, made it impossible to derive a reliable line profile from the spectra \citep{Anitra2023}.}. 

We fitted the eight spectra simultaneously using a power-law model to account for the normalised continuum (the associated parameters are not physically relevant), to which we added the \textsc{diskline} model to fit the H$\beta$  line profile. 
The observations were obtained within a time interval of 2.75 h (one orbital period), and we therefore expect very little spectral variation in the different spectra. As a result, we constrained specific parameters, such as the inclination angle, emissivity index, and inner and outer radii, to be the same for all the spectra.
The model, called \textsc{model 1}, achieved a poor fit to the data, with a $\rm{\chi^{2}/d.o.f.}$= 4089.1/2811. Nonetheless, the values for the inner and outer radii of the line emission region within the disc, $(9.6 \pm 0.2) \times 10^{3} \, \rm{R_{g}}$ ($\sim 0.16\, R_{T}$) and $5.4^{+0.1}_{-0.2}\, \times 10^{4} \, \rm{R_{g}}$ ($\sim 0.92\, R_{T}$), respectively, agree with expectations. Additionally, the emissivity index characterises a region of the disc that is located farther from the compact object.
However, as shown in Fig. \ref{j1357_disk} (second panel), residuals lie in the region between the two peaks. 

To improve the fit without modifying the \textsc{diskline} model, we attempted two different approaches. First, we masked out the core of the line and excluded these data from the fit. The resulting best-fit parameters are consistent with those presented previously, with a $\rm{\chi^{2}/dof}=\,4071.72/2805$.
Second, we introduced a Gaussian absorption line for which the centroid and FWHM parameters were linked for all the spectra, but we allowed an independent variable depth to determine whether each spectrum significantly required this feature (see Sect. \ref{discussion}).
This model, called \textsc{model 2}, achieved a better fit to the data with a $\rm{\chi^{2}/dof}=\,3849.0/2794$. We tested the improvement of the fit using the statistical test \textsc{F-test} and obtained a probability of a chance improvement of $\sim 1.1 \times 10^{-28}$. This means that including the Gaussian line in absorption improves the quality of the fit with a confidence level (c.l.) higher than 7$\sigma$.
The new model provides a more solid constraint on the inclination angle (i=83$^{+5}_{-3}$ degrees), and the H$\beta$ emission line is consistent with being emitted by a ring in the disc with a radius between $9.6^{+0.2}_{-0.1}\times 10^{3}\, \rm{R_{g}}$ ($\sim 0.16\, R_{T}$) and $ 5.7^{+0.1}_{-0.2} \times 10^{4}\, \rm{R_{g}}$ ($\sim 0.91\, R_{T}$).
{However, for this high number of degrees of freedom, a reduced $\chi^{2}$  of 1.38 is unacceptable because it
implies a very low null-hypothesis probability.
In order to obtain reliable error bars for the best-fit spectral parameters, we therefore scaled the errors by the $\sqrt{\chi_{red}^{2}}$ to achieve a reduced $\chi^{2}$ of approximately 1 and repeated the fit.
Using this approach, we obtained the same values for the inner and outer radii of the emitting region, while the fit was able to provide just a lower limit for the inclination angle of 80 degrees. Notably, this finding only holds true for errors calculated at the 90\% c.l. Conversely, when errors are calculated at the 68\% c.l, the inclination angle is constrained to be $83.8^{+5.4}_{-1.8}$ degrees (see Table \ref{tab_diskline}).
\begin{figure}
    \centering
    \includegraphics[width=.5\textwidth]{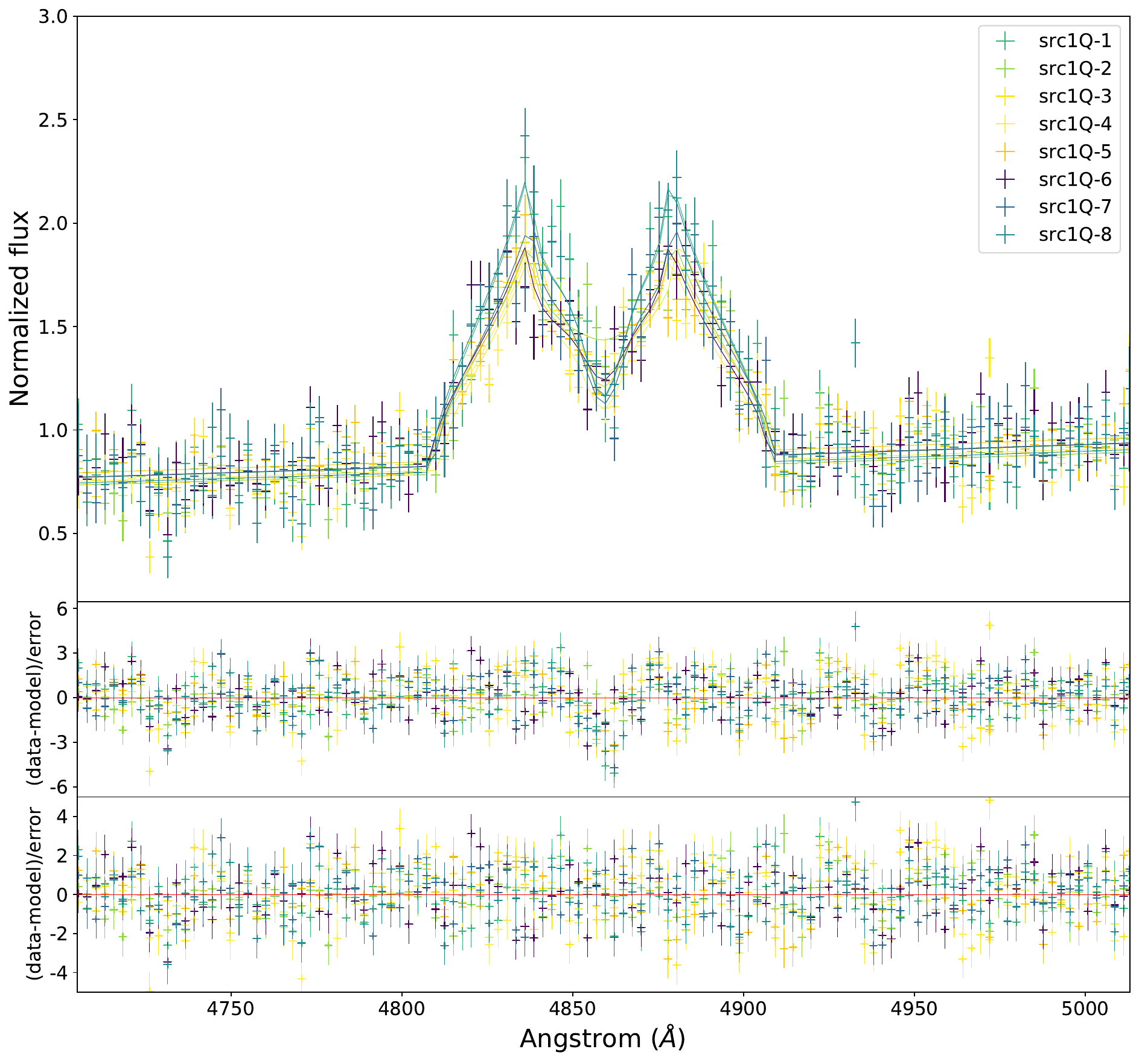}
    \caption{J1357 spectra collected during a quiescent phase and residuals in units of sigma with respect to the \textsc{diskline} model described in the test (top panel). The middle and bottom panel shows the residuals obtained by using only the \textsc{diskline} model (\textsc{model 1}) and those obtained with the same model plus a Gaussian absorption line (\textsc{model 2}), respectively.}\label{j1357_disk}
\end{figure}

\subsection{J1305 spectroscopy}\label{j1305_analysis}
The wavelength coverage and spectral resolution provided by the telescope set-up allowed us to perform a high-resolution analysis focused on the H$\alpha$ emission line.
In the spectra collected during the outburst, we identify H$\alpha$ and H$\beta$ emission lines, corresponding to the Balmer series and He II 4686 \AA \, as the Bowen blend (a mixture of \ion{N}{iii} and \ion{C}{iii} emission lines; see e.g. \citealt{Steeghs&Casares2002}).
We decided to focus our analysis on the H$\alpha$ line, which is present in all the spectra in the two days and is the strongest isolated feature \citep{Miceli2023_submitted}.
We separately analysed the two days of optical observations. For the first day, we used four of the six spectra 
excluding, in particular, src2-O1 because of its strong absorption component and src2-O3 because it does not show a double-peak profile, but a flat top profile (which recalls outflow features; \cite{Miceli2023_submitted}, see also \citealt{Cuneo2020}). 

We applied \textsc{model 1} to the data and linked the parameters of each spectrum as described in the previous section (see Fig. \ref{fig:day1_diskline_fit}).
On the second day of observations, a distinct absorption component is evident in all spectra. It is located at wavelengths longer than those associated with the \ha{} line (about 6563 \AA) (see Fig. \ref{fig:diskline_fit2}). 
To take this feature into account, we added to the previous model a Gaussian absorption line for which the centroid and FWHM parameters we linked for all the spectra, but we allowed an independently variable depth.
For both days, the adopted models achieve a reasonably good fit to the data, with a $\rm{\chi^{2}/dof}=\,139.4/153$ (first day) and $\rm{\chi^{2}/dof}=\,149.0/222$ (second day). The best-fit parameters are reported in Table \ref{tab_diskline}. 
On both days, the inclination angle is $70\pm4$ and $71 \pm 4$ degrees, which is consistent with the angle reported in the literature ($i=72_{-8}^{+5} \, \rm{degrees}$; \citealt{2021_mata-sanchez}). Furthermore, for the first day, we derived inner and outer radii values of $\rm{1.20^{+0.09}_{-0.07}\times 10^{5}\, \rm{R{g}}}$ ($\sim 0.83\, R_{T}$) and $\rm{1.96^{+0.32}_{-0.27}\times 10^{6}\, \rm{R{g}}}$ ($\sim 13\, R_{T}$), respectively. For the second day, these values were $\rm{1.66^{+0.10}_{-0.08}\times 10^{5}\, \rm{R{g}}}$ ($\sim 1.1\, R_{T}$) and $\rm{2.73^{+0.9}_{-0.6}\times 10^{6}\, \rm{R{g}}}$ ($\sim 19\, R_{T}$) . A more detailed discussion of these parameters is presented in the following section.

\subsubsection{Quiescence}
We chose 10 of the 16 spectra for the quiescent phase 
 that showed a more regular line profiles. Our preference lay with line profiles with a symmetrical double peak because the asymmetry in the intensity of the two peaks can be caused by various effects, for instance a hot spot \citep{shafter_1986}, and this might complicate the fitting. However, the exclusion of the 6 spectra does not compromise the credibility of the model. As a test, we individually analysed the excluded spectra and obtained parameter values that were entirely consistent with those reported in the main analysis (see Sect. \ref{degenerazione}). The symmetry of a profile can be estimated by examining the ratios of the two peaks, as reported in Table \ref{ratio}.
\begin{table}[]
\renewcommand{\arraystretch}{1.25} 
    \centering
    \caption{\label{ratio} Ratios of the blue and red peaks of the J1305 spectra collected during quiescence.}
    \begin{threeparttable}
    \begin{tabular}{lc|lc}
    \hline
    \textsc{spectra\tnote{*}} & \textsc{$\rm{I_{B}/I_{R}}$} &\textsc{spectra} & \textsc{$\rm{I_{B}/I_{R}}$} \\
    \hline
\textbf{src2-Q1} & $ 1.03 \pm  0.09$ & \textbf{src2-Q9 }& $ 1.05 \pm  0.06$ \\
\textbf{src2-Q2} & $ 0.97 \pm  0.09$ & \textbf{src2-Q10} & $ 1.20 \pm  0.07$ \\

\textbf{src2-Q3} & $ 1.18 \pm  0.06$ & src2-Q11 & $ 1.34 \pm  0.07$ \\

\textbf{src2-Q4} & $ 1.17 \pm  0.08$ & src2-Q12 & $ 1.28 \pm  0.07$ \\

\textbf{src2-Q5 }& $ 1.13 \pm  0.06$ & src2-Q13 & $ 1.64 \pm  0.09$ \\

\textbf{src2-Q6} & $ 1.22 \pm  0.06$ &src2-Q14 & $ 1.51 \pm  0.09$ \\

src2-Q7 & $ 1.55 \pm  0.06$ & \textbf{src2-Q15} & $ 0.77 \pm  0.08$ \\

src2-Q8 & $ 1.41 \pm  0.07$ & \textbf{src2-Q16} & $ 0.77 \pm  0.11$ \\

\hline
    \end{tabular}
    \begin{tablenotes}
    \item[*] The bold values represent the spectra chosen for the spectral analysis.
\end{tablenotes}
\end{threeparttable}

\end{table}
Following the same procedure as applied to J1357, we first fitted the spectra applying \textsc{model 1}. However, the latter did not reach a good fit to the data, with a $\chi^{2}$/dof of 6432.32/2650, because residuals were clearly present in the core region (see the first panel of Fig. \ref{fig:out_diskline_fit2}).
Therefore, we applied \textsc{model 2} and linked the parameters of different spectra as described for the J1357 spectra. The model achieved a better fit to the data, ensuring a $\chi^{2}$/dof of 4161.1/2620. The $\chi^{2}$ value had to be rejected here as well. Therefore, as discussed in the previous section, we multiplied the errors by the square root of the reduced $\chi^{2}$ and repeated the fitting.
We obtain a constraint on the inclination angle of $72.6^{+1.4}_{-1.3}$ degrees at 90\% c.l., as well as best-fit values for the inner and outer radii of the emitting region associated with the \ha{} emission line of $1.94^{+0.07}_{-0.09} \times  10^{4}\, \rm{R_{g}}$ ($\sim 0.1\, R_{T}$) and $2.01 \pm 0.04 \times 10^{5}\, \rm{R_{g}}$ ($\sim 1.3\, R_{T}$), respectively.
\begin{figure}
    \centering
    \includegraphics[width=.5\textwidth]{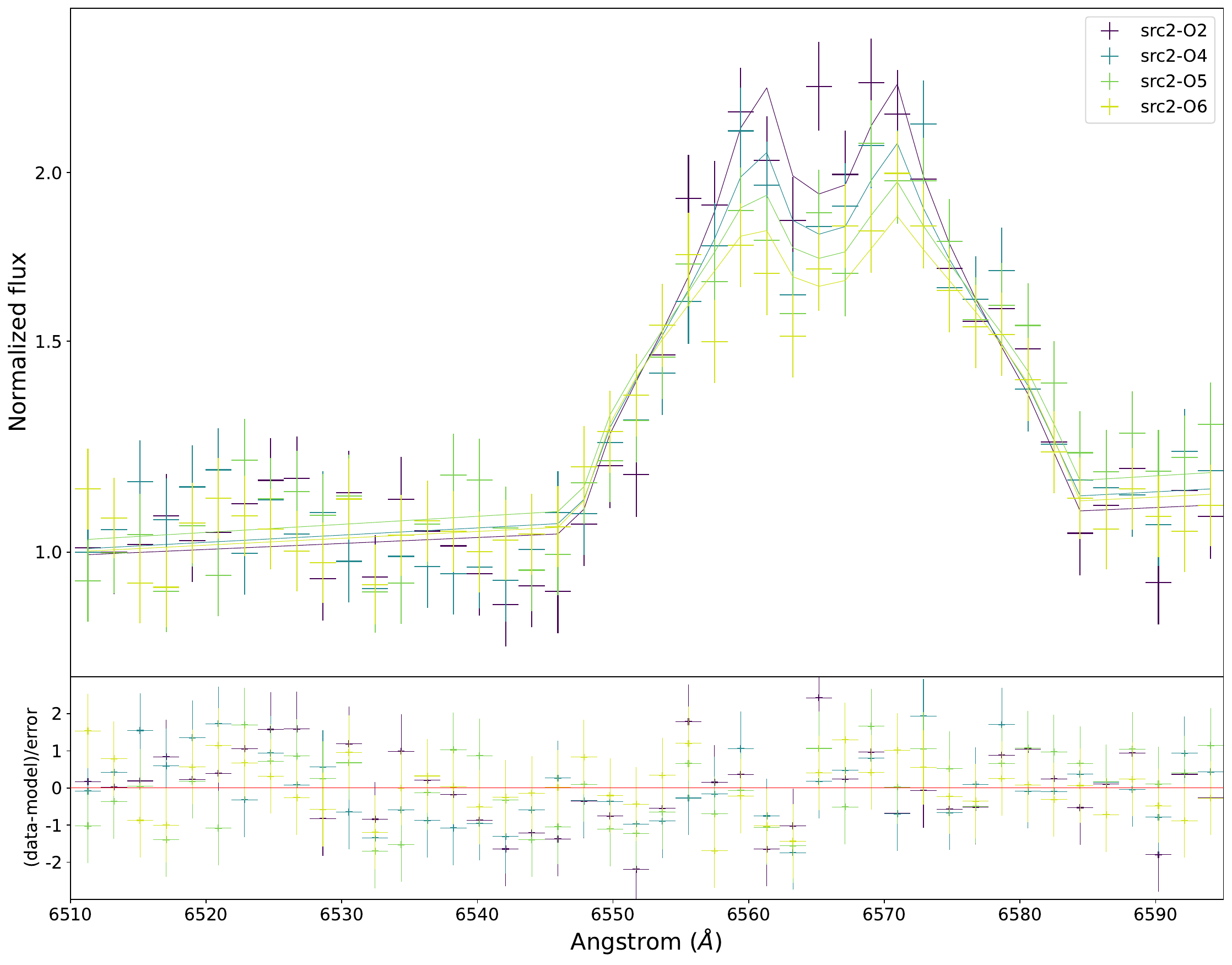}
    \caption{J1305 spectra collected during an outburst phase and residuals in units of sigma with respect to the \textsc{diskline} (\textsc{model 1}), related to the first day of observation.}\label{fig:day1_diskline_fit}
\end{figure}
\begin{figure}
    \centering
    \includegraphics[width=.5\textwidth]{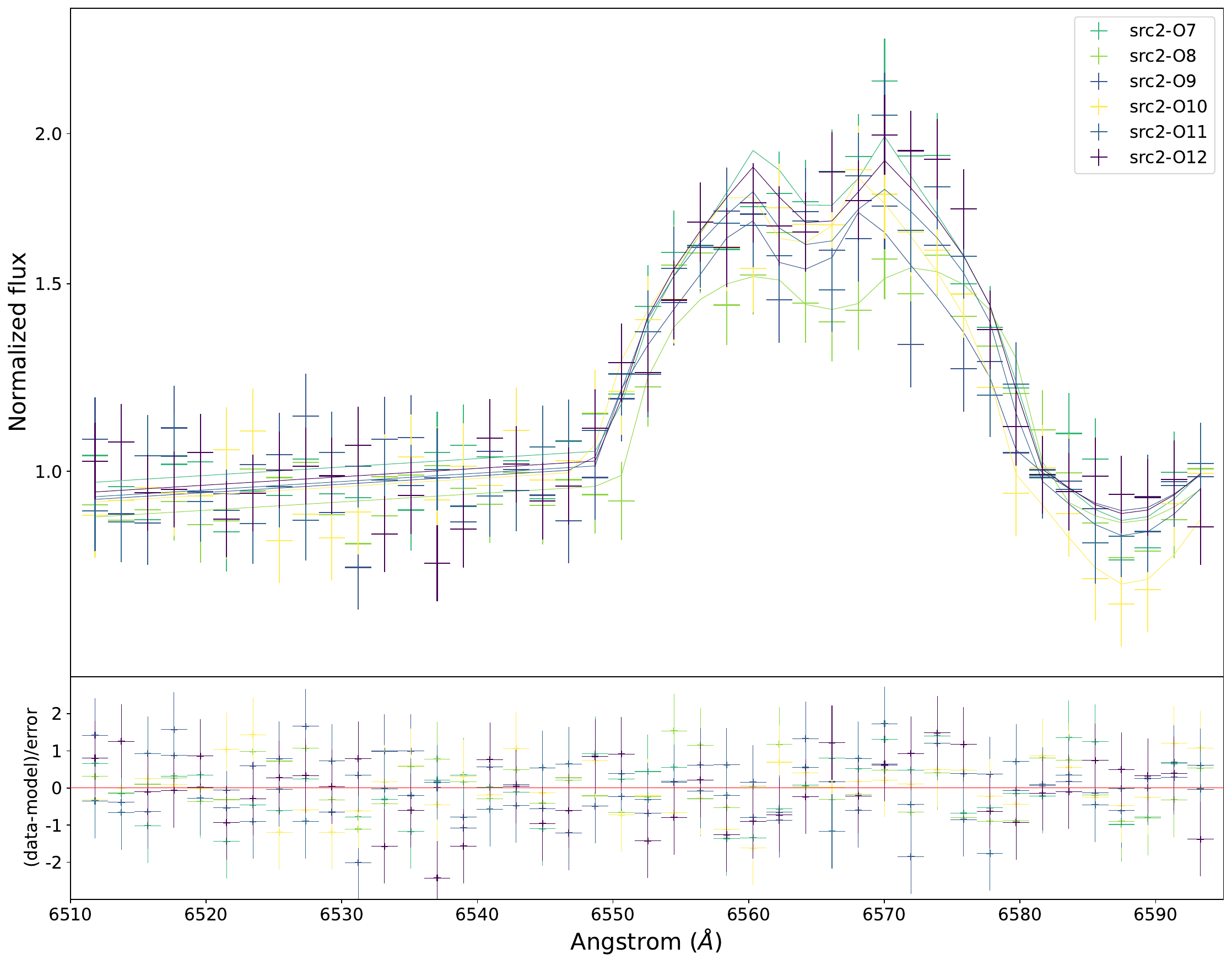}
    \caption{J1305 spectra collected during an outburst phase and residuals in units of sigma with respect to the \textsc{diskline} model plus an absorption Gaussian line, related to the second day of observation.}\label{fig:diskline_fit2}
\end{figure}
\begin{figure}
    \centering \includegraphics[width=.5\textwidth]{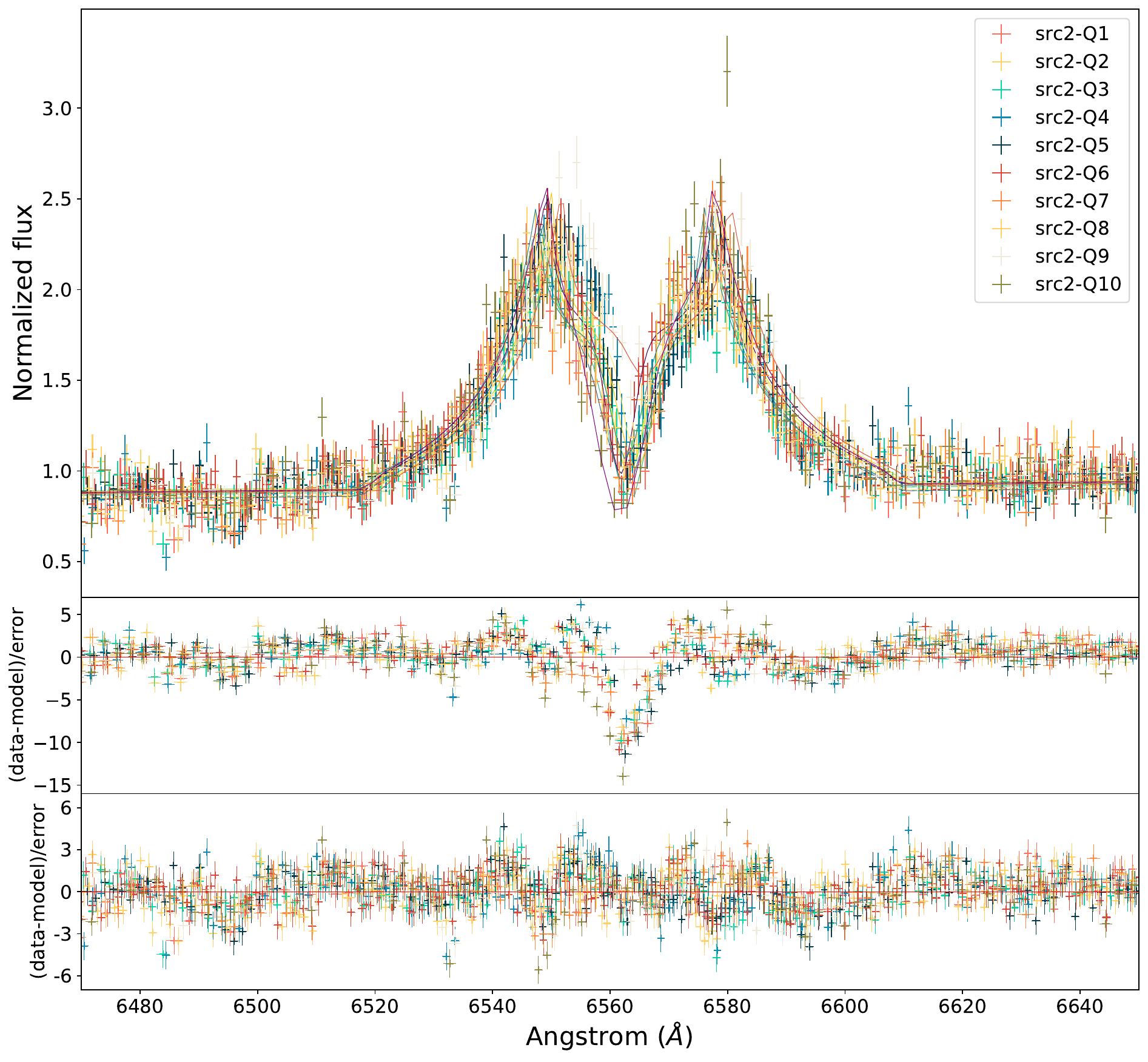}
    \caption{J1305 spectra collected during a quiescent phase and residuals in units of sigma with respect to the \textsc{diskline} model described in the text (top panel). The middle and bottom panels show the residuals obtained with the \textsc{diskline} model alone (\textsc{model 1}) and those obtained with the same model plus a Gaussian absorption line (\textsc{model 2}).}\label{fig:out_diskline_fit2}
\end{figure}

\subsection{\label{degenerazione}Analysis of the parameter degeneration}
The line profiles can easily be affected by external factors, such as hot spots or disc precession (e.g.\cite{Orosz_1994,Mason_2000MNRAS.318..440M,Marsh_Robinson_1994MNRAS.266..137M}). 
These potential pitfalls for a model might prompt concerns about the reliability of the approach. We therefore explore degeneracies and systematics quantitatively here to ensure that the previously derived constraints are statistically reliable.

When multiple parameter combinations yield similar or indistinguishable results during the model-fitting process, this indicates that the best-fit values of the individual parameters cannot be determined based on the data alone. To investigate this degeneracy, we employed the \textsc{steppar} command in \textsc{xspec} to perform a fit while stepping the value of the parameters through a given range. This approach allowed us to evaluate all possible pairings of the parameters of interest, measuring how the $\chi^{2}$ value changes in relation to the best fit for each combination. This allowed us to identify any value combinations that might result in an equivalent spectral fit.
We performed the \textsc{steppar} on data sets collected during  quiescence from the two
sources and focussed on parameters such as inclination, outer radius, and inner radius, which significantly affect the line profile. In Fig. \ref{steppar}, we present contour plots obtained from the analysis.

The contours of the regions are clearly defined and concentric around the best-fit value, indicating robustness in the parameter estimates. In addition, the contours lack extended overlapping or elongated areas that stretch along the diagonals, which would typically indicate degeneration. It is important to observe that even in regions with a diagonal trend (panels a and d), the contours delineate parameter values that lie within the error limits reported in the text. This shows that the same $\chi^{2}$ values cannot be achieved with parameter pairs that deviate from those established by the best fits.

Because the physics within an accretion disc is complex and the approach of the \textsc{diskline} model is simplistic, the high precision of the obtained inclination constraints (especially from the quiescence spectrum of J1305) may cast doubt on the reliability of the results.
An insightful way to re-evaluating and discussing the error associated with the best-fit parameters, especially the inclination angle, involves fitting each spectrum individually and considering the dispersion of the resulting values.

We therefore fitted the spectra acquired during quiescence for J1357 and J1305 separately, allowing all parameters to vary freely\footnote{We chose not to fit the outburst spectra because the derived inner and outer radii are not very reliable}. In Table \ref{inclination} we list the inclination values for each fit. Moreover, we decided to also include the spectra that were rejected in the analysis presented in the text to avoid any bias in the estimation of the inclination angle error. We ensured through a \textsc{steppar} analysis throughout the entire range of possible inclination values that the fit did not yield a value corresponding to a local minimum of the $\chi^{2}$. Subsequently, we determined the best fit as the average of the obtained values, with the associated error calculated as the semi-dispersion because it represents the maximum dispersion around the average value. This yielded an inclination of 81 $\pm$ 5 degrees for J1357 and 73 $\pm$ 4 degrees for J1305, and these inclinations agree with the values derived from the analysis. Moreover, the error bars obtained with this method are consistently similar to those constrained by the \textsc{diskline} model.
\begin{table}[]
    \centering
    \caption{\label{inclination}Inclination values obtained by separately fitting the spectra related to the two sources in quiescence. }
\begin{threeparttable}
\begin{tabular}{cccc}
\hline
J1305 &  Inc (degrees)  & J1357 & Inc (degrees)\\ \hline
src2-Q1 & 74.1 &  src1-Q1 & 82.4 \\ 
src2-Q2 & 71.5 &  src1-Q2 & 83.3 \\ 
src2-Q3 & 73.1 &  src1-Q3 & 82.7 \\
src2-Q4 & 72.7 &  src1-Q4 & 86 \\ 
src2-Q5 & 72.3 &  src1-Q5 & 83.3 \\ 
src2-Q6 & 73.2 & src1-Q6 & 75.7 \\ 
src2-Q9 & 76.6 & src1-Q7 & 79.3 \\ 
src2-Q10 & 74.1  & src1-Q8 & 75.3 \\ 
src2-Q15 & 73.6  &  & \\
src2-Q16 & 74.5  &  & \\ 
src2-Q7& 69.7& & \\
src2-Q8& 69.4& & \\
src2-Q11& 76.0& & \\
src2-Q12& 70.7& & \\
src2-Q13& 71.5& & \\
src2-Q14& 72.5& & \\
\hline
 Best value\tnote{*}& $73 \pm 4 $ & &$81\pm 5$ \\
\hline
\end{tabular}
\begin{tablenotes}
    \item[*]The best value was obtained by averaging the values, and the associated error is the semi-dispersion of the values.
\end{tablenotes}
\end{threeparttable}

\end{table}
\\ \\
\begin{figure*}
    \centering
    \includegraphics[width=.45\textwidth]{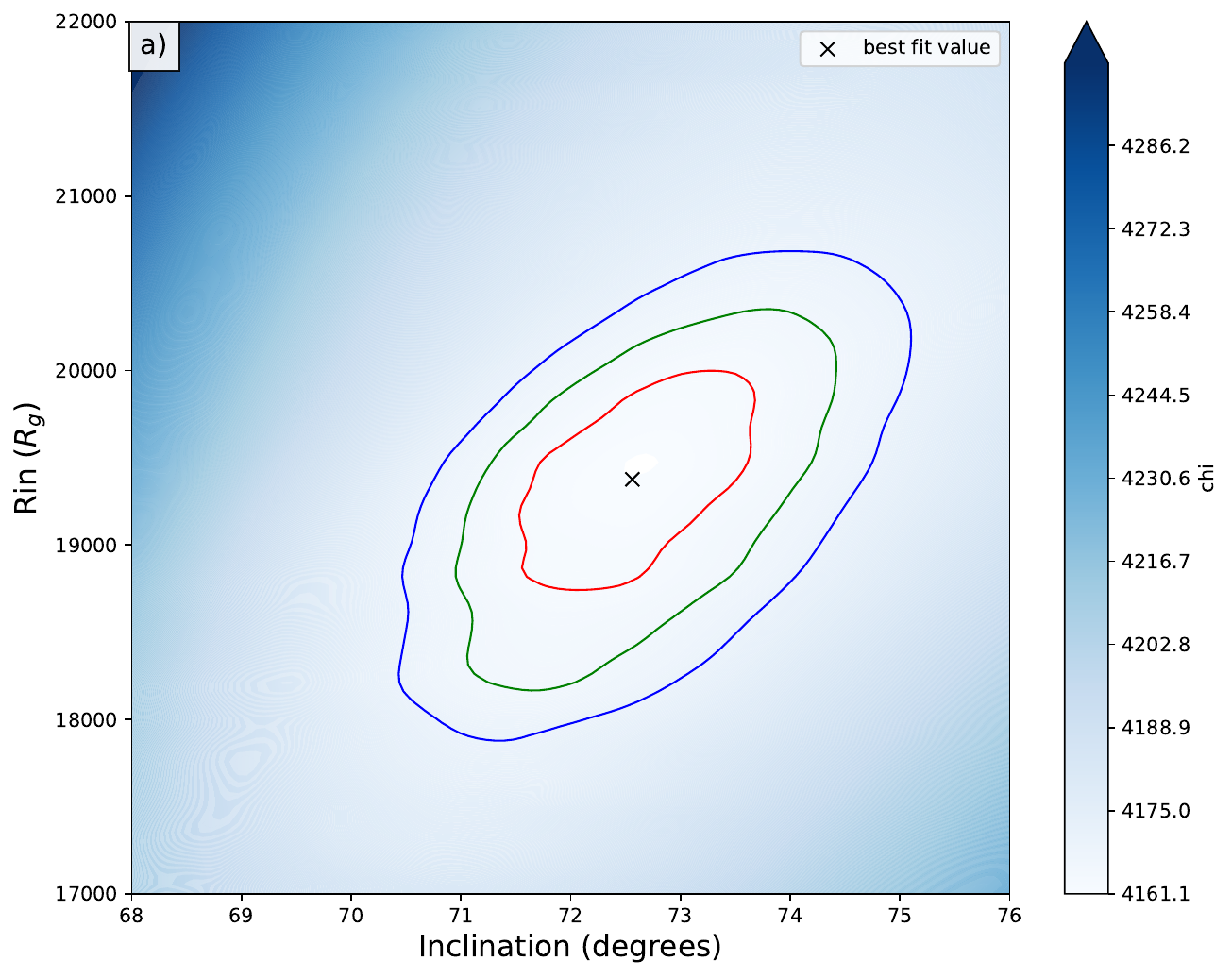}\quad
    \includegraphics[width=.45\textwidth]{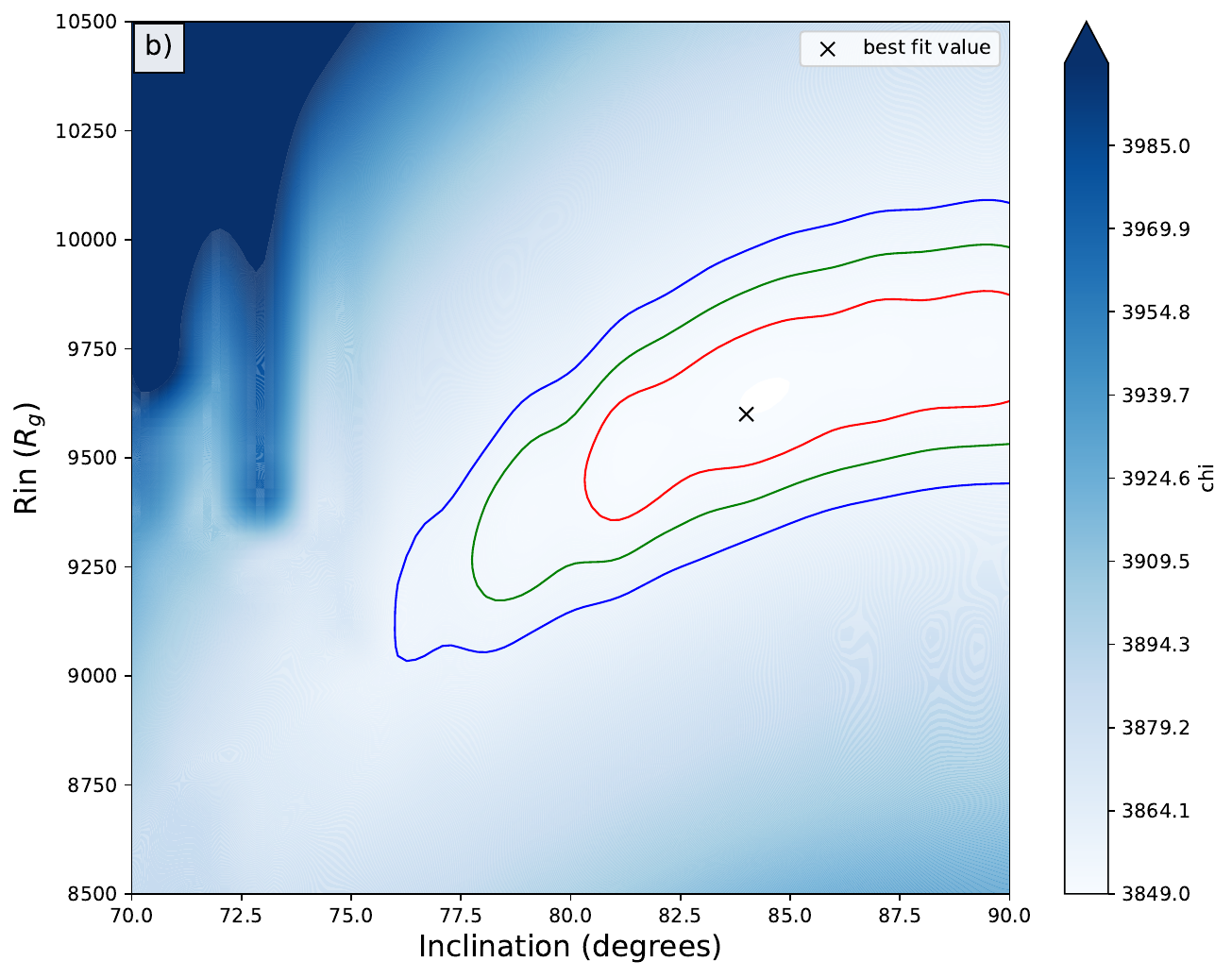}\quad
    \includegraphics[width=.45\textwidth]{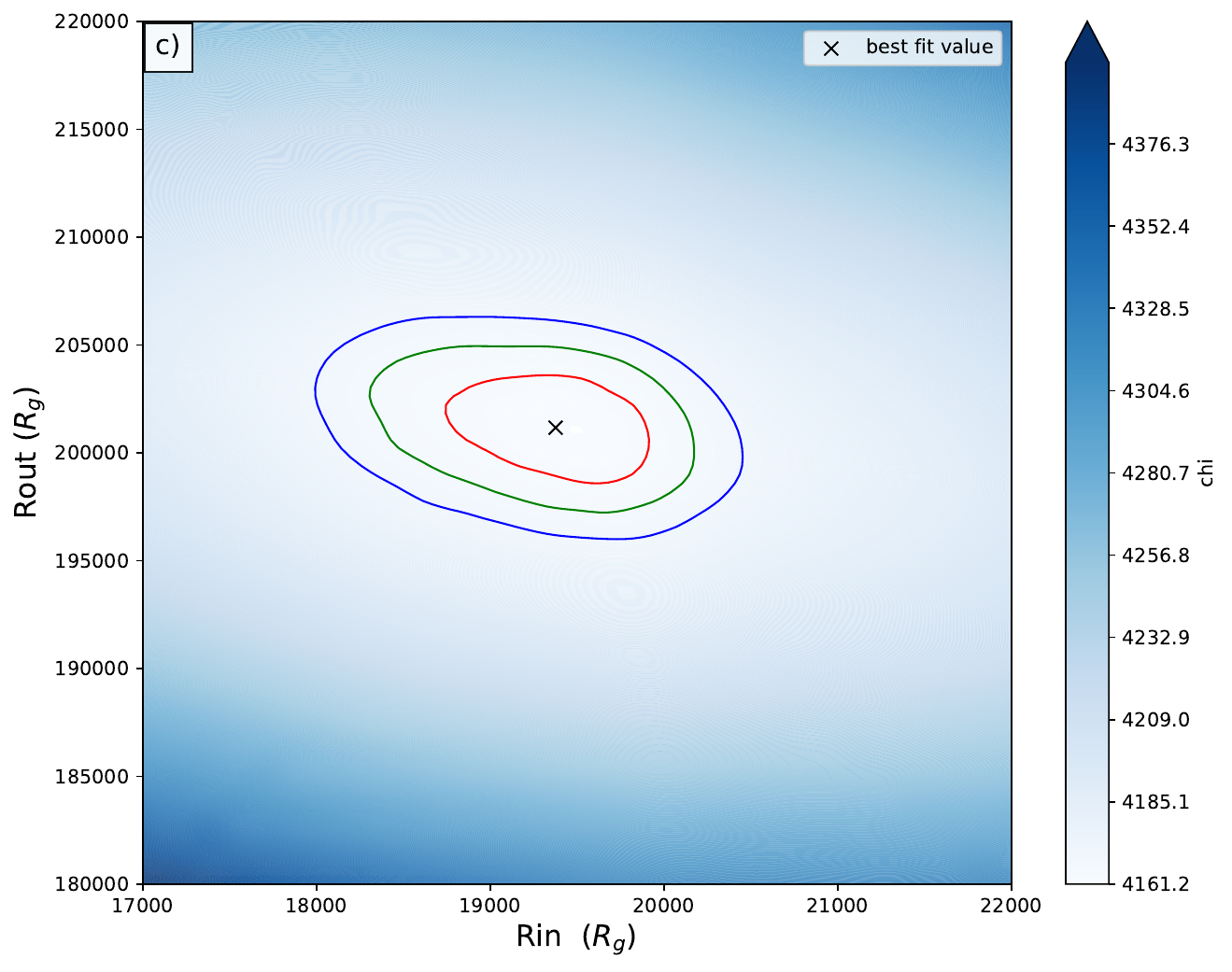}\quad
    \includegraphics[width=.45\textwidth]{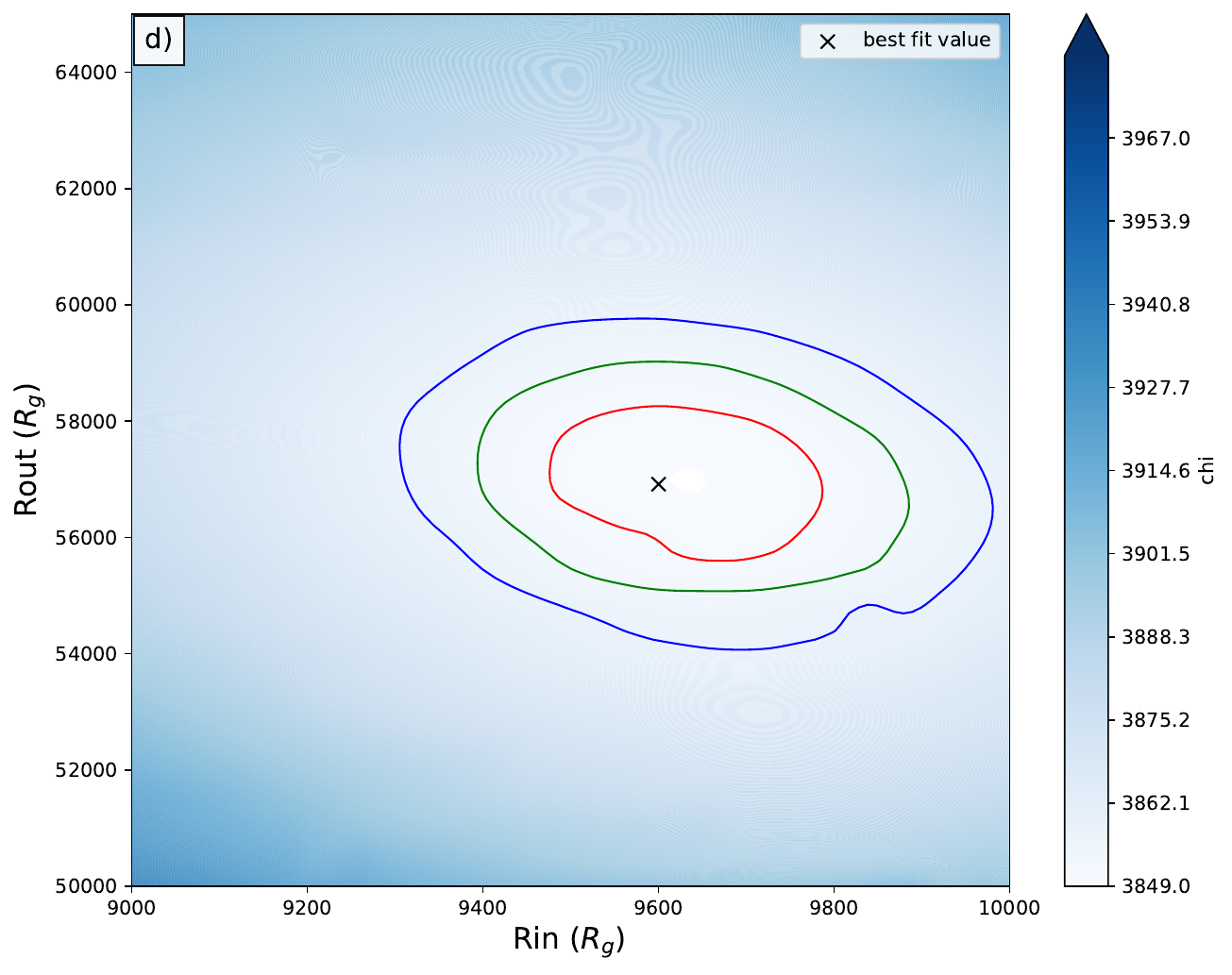}\quad
   \includegraphics[width=.45\textwidth]{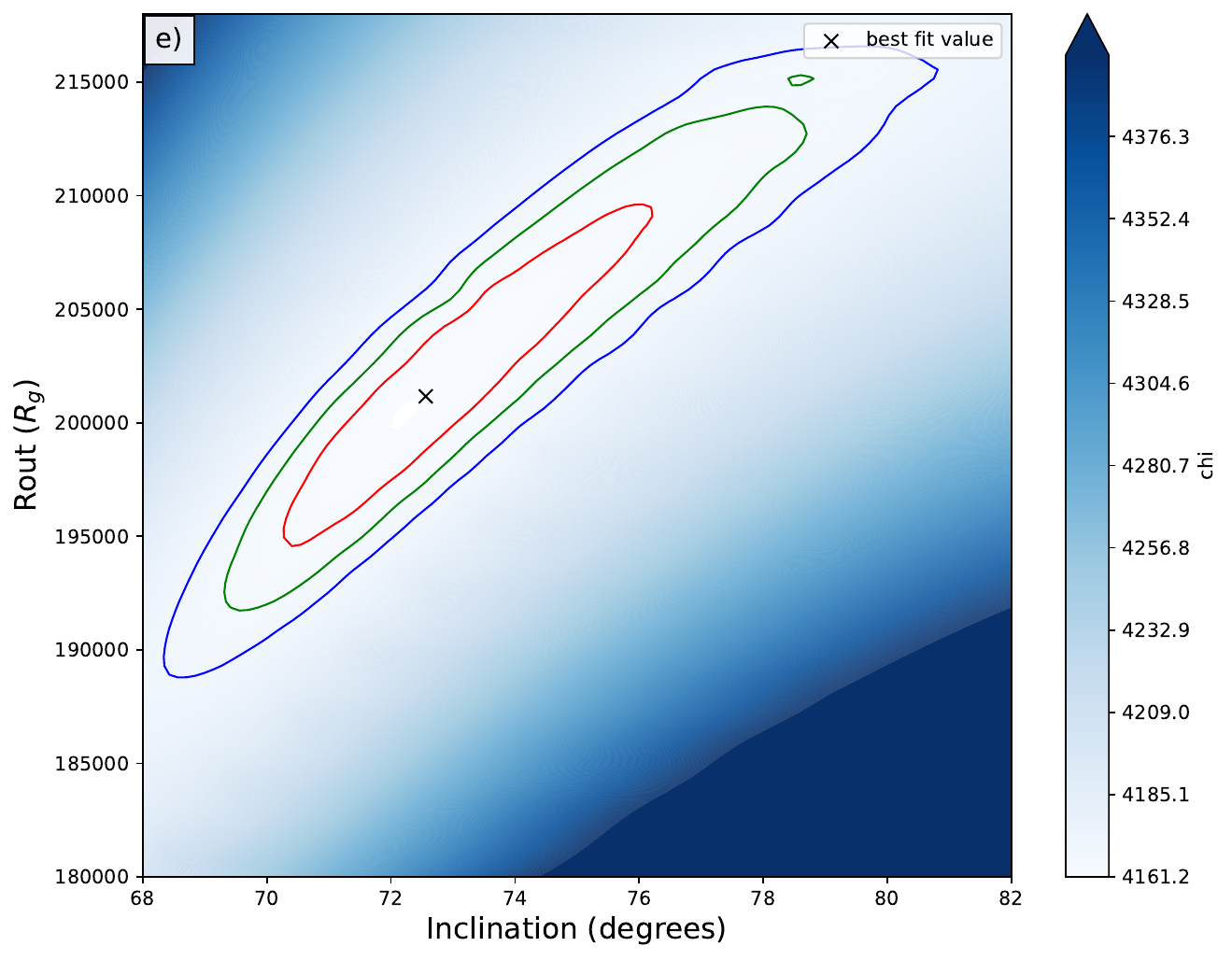}\quad
    \includegraphics[width=.45\textwidth]{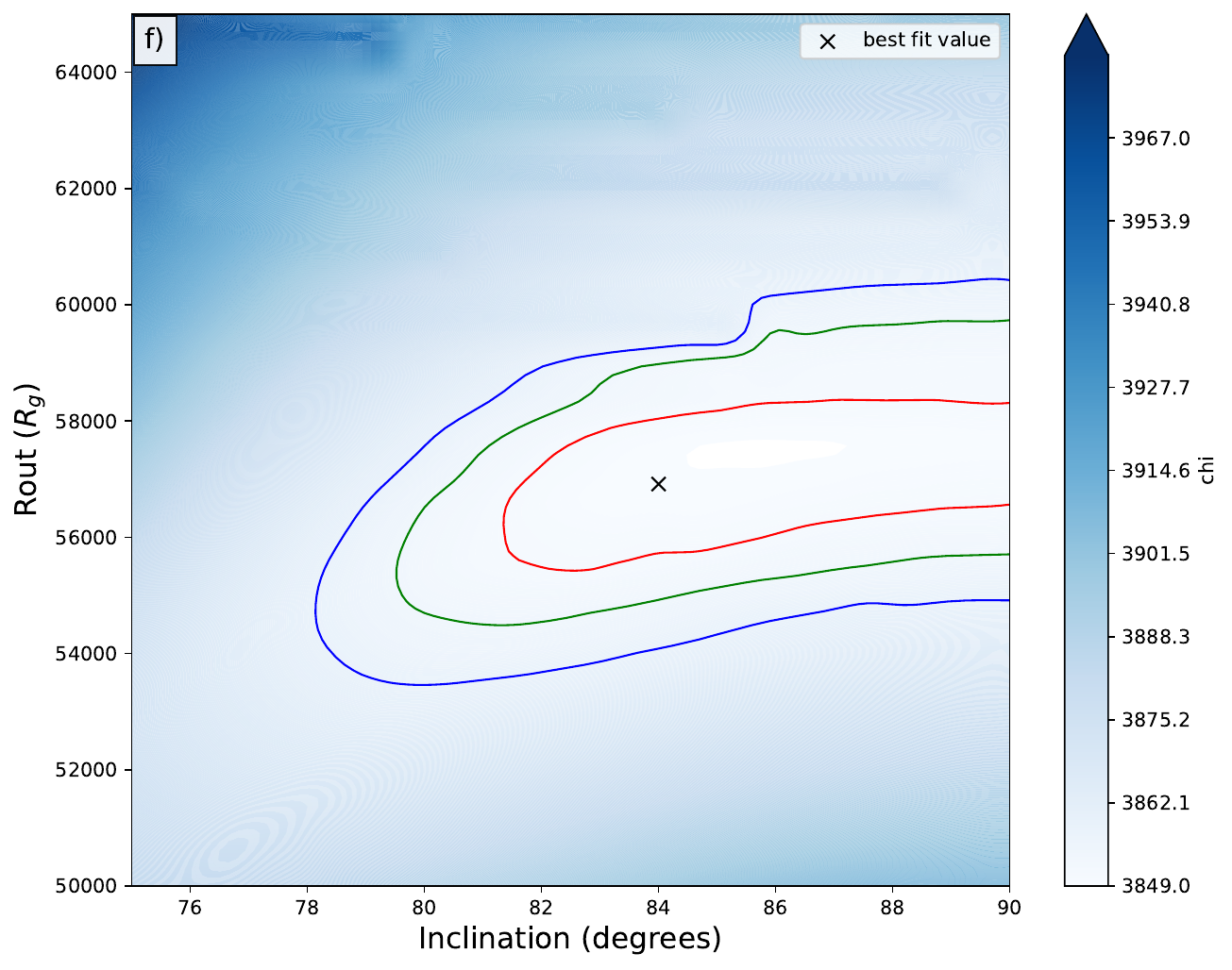}
    \caption{\label{steppar}Contour plots for inclination and the outer and inner radii of the spectra related to J1357 (first column) and J1305 (second column). The contours represent the 1$\sigma$, 2$\sigma$, and 3$\sigma$ confidence levels, and the cross marks the best-fit values obtained from the best fit.}
\end{figure*}

\section{Discussion}\label{discussion}
We analysed GTC observations of J1357 collected during quiescence, along with VLT and \textit{Magellan} observations of J1305, taken during quiescence and outburst phases, respectively. Our study focused on the double-peaked profiles of the \hb{} and \ha{} emission lines. With the aim of gaining insights into the system geometry, we employed the \textsc{diskline} model, which is complementary to traditional analysis approaches. 
As discussed in the previous section, the model including the \textsc{diskline} component along with a (Gaussian) absorption line 
provides a good fit to the data of both systems. This significantly improved the quality of the fit with respect to the model without the absorption line.

The addition of this feature may be necessary due to the potential limitations in the assumptions made by the  \textsc{diskline} model. 
It describes a Gaussian emission line profile, modified by the Doppler effect induced by a Keplerian velocity distribution, in an 
accretion disc.
It is clear that the model does not take self-absorption or other features due to the optically thick nature of the disc into account, which was also noted by \citet{Orosz_1994}.
This effect is particularly strong for high-inclination angles, as appears to be the case of J1357 and J1305.
This agrees with our results, for which an additional Gaussian absorption enabled a better description of the observed spectra. 

Upon visual examination of the individual spectra, it becomes clear that the central core of the line displays narrow and variable absorption that changes in depth and even reaches (sometimes going below) the continuum average level. 
Features like this 
were mainly detected in cataclysmic variables that were observed at a high-inclination angle (with $i\, > 75$ degrees, \citealt{1983_Schoembs}), and they are thought to be caused by occultation of the inner regions of the accretion disc.
Differently from the effects caused by self-absorption, narrow cores are variable with the orbital phase, and they can reach much deeper than the optically thick absorption effects. They can reach even below the normalised continuum \citep{1981_rayne_whelan}.
Narrow variable cores in the J1357 spectra were first claimed by \citet{Daniel_2015}, who noted a variation in the normalised flux of the \ha{} core. This was later confirmed by \citet{Anitra2023}, who analysed high-resolution data focused on the H$\beta$ emission line.  
\cite{2021_mata-sanchez} performed dynamical studies of the same VLT data of J1305 as we presented here. They reported narrow cores in the spectra. 

As further proof of this, we calculated the significance of the absorption feature in each spectrum in units of sigma by comparing the normalisation of the line with its uncertainty at the 68\% c.l. 
We found that the intensity of the line in J1357 src1-Q2, src1-Q3, and src1-Q6 is $\leq 3 \sigma$, meaning that during these orbital phases, the absorption core of the line is negligible, but it is stronger in  src1-Q1, src1-Q7 and src1-Q8. 
In the quiescent spectra of J1305, all the absorption lines are statistically significant at more than  $7 \sigma$, except for the line associated with src2-Q15, where this feature is not necessary.
This result is very similar to the result reported by \citet{MArsh_1987} in the dwarf nova system Z Cha, which exhibited a core depth variability along the orbit in every line of the Balmer series in the spectrum.

As reported in Sect. \ref{j1305_analysis}, spectra collected during the outburst do not show narrow variable cores. However, the spectra collected on the second day of observation display absorption features at higher wavelengths compared to the \hb{} line, which may resemble an inverted P-Cygni profile, which is associated with inflows \citep{Cuneo2020}. \cite{Miceli2023_submitted} have extensively discussed the nature of this phenomenon and proposed that a broad and variable absorption component is observed in all spectra, which can affect the shape of the line. Nevertheless, the origin of this feature remains the subject of ongoing debate.

\subsection{Geometry of the emitting region}
\subsubsection{Inclination angle}
The best-fit parameters of both sources provide a description of the geometry of the system in line with the hypothesis of a high inclination. 
We obtain an angle of $72.6^{+1.4}_{-1.3}$ degrees ($73 \pm 4$ degrees considering the semi-dispersion around the average value)  for J1305 during quiescence and angles of $70 \pm 4$ and $71 \pm 4$ degrees for the two days of outburst, respectively. These results are consistent with the previous constraint reported by \citet{2021_mata-sanchez}, $i=72^{+5}_{-8}$ degrees.

Recently, \cite{Casares_2022} studied the inner core depth in J1357 spectra and derived an inclination angle of $87.4^{+2.6}_{-5.6}$ degrees, which matches the estimate we find by applying the \textsc{diskline} model (i>80 degrees at 90\% c.l., $i=80^{+2}_{-6}$ degrees at 68 \% c.l. and $81 \pm 5$ considering the average values). However, the lack of eclipses in both the X-ray and optical light curves of this source \citep{Corral_santana_2013} contradicts the expected behaviour of an edge-on configuration. \citet{Corral_santana_2013} provided an explanation for this phenomenon, which may be due to the low mass ratio of the system, implying that the radius of the donor star is either comparable to or smaller than that of the outer rim of the disc. Consequently, even in an edge-on configuration, the central disc region is not obstructed by eclipses of the companion star.
It is possible to give quantitative credibility to this hypothesis by evaluating whether the mass ratio $q$ given in the literature allows the occurrence or absence of eclipses at a particular inclination angle. 
\begin{figure}
    \centering
    \includegraphics[width=.5\textwidth]{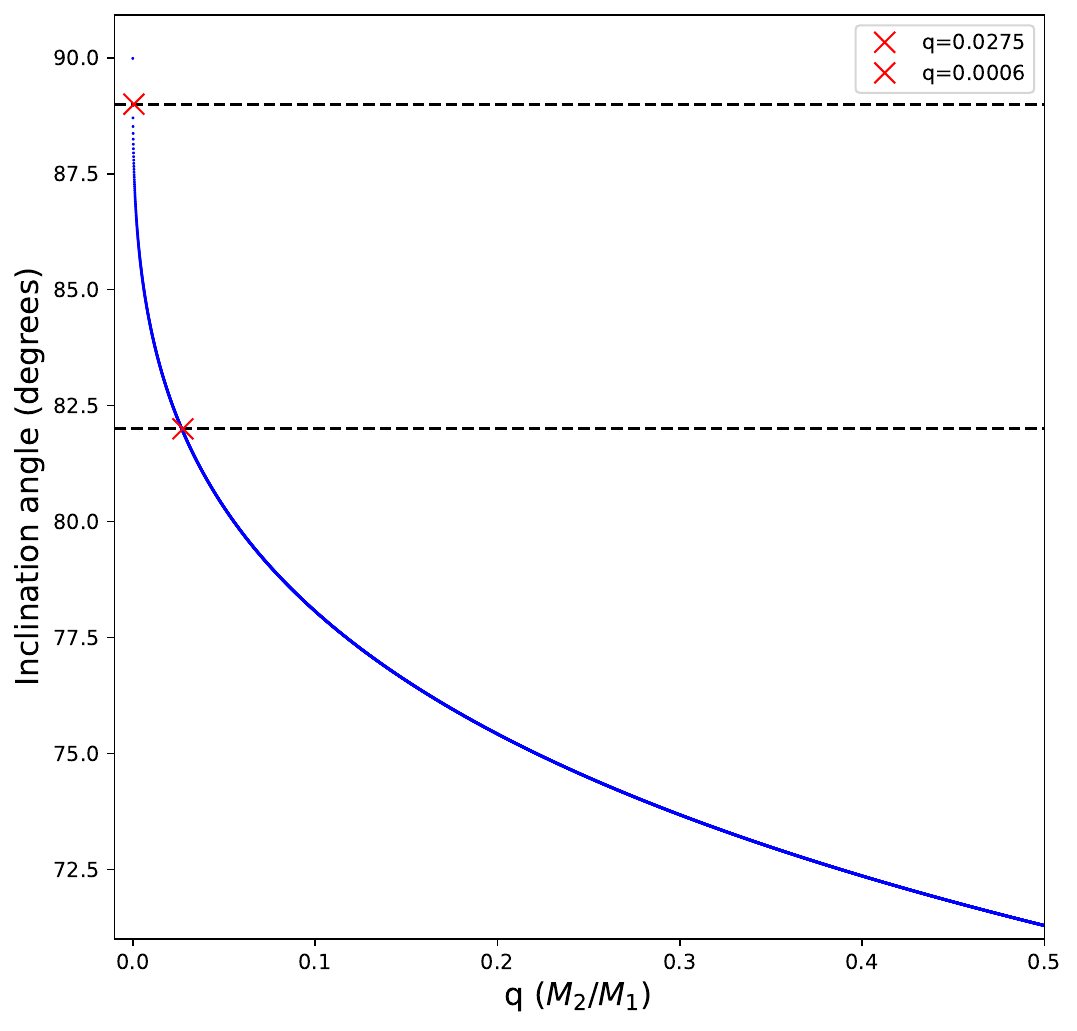}
        \caption{\label{inclinationVSq} Inclination angle obtained with Eq. \ref{theta} by varying the mass ratio $q$ in a range of 0 to 0.5.  The dashed black lines represent the lower and upper bounds of the inclination values determined for J1357 with \textsc{model 2}. Consequently, the data points highlighted with a red cross symbolise the lower and upper limits of the mass ratio $q$ allowed for the system in order to preserve the lack of eclipses at this specific inclination.}
\end{figure}
As stated by \cite{2018_iaria},  the angle $\theta$ between the line of sight and the equatorial plane can be described by the following equation:
\begin{equation}\label{theta}
    \tan \theta=\left[\frac{R_2^2-x^2}{a^2-\left(R_2^2-x^2\right)}\right]^{1 / 2} ,
\end{equation}
where $R_{2}$ is the Roche-lobe radius of the companion star, 
and $x$ is the obscured region during the eclipse.
Without an eclipse, $x$ can be assumed to be zero, while the Roche-lobe radius can be calculated using Eq. \ref{roche} by substituting $q^{-1}$ with $q$.
By substituting $R_{2}$ into the previous equation, the angle $\theta$ is clearly only a function of the mass ratio $q$. 
In order to investigate the combinations of $q$ and $\theta$ that preserve the condition without an eclipse, we varied the $q$ parameter within the range of 0 to 0.5 and determined the corresponding values of $\theta$. The results are shown in Fig. \ref{inclinationVSq}.

For an inclination of $83.8^{+5.4}_{-1.8}$ degrees obtained at  68\% c.l, we can establish a $q$ ratio range between 0.0006 and 0.0275, which is consistent with the value reported by \citet{Daniel_2015}. Moreover, with the mass function for $M_{1}$, $\rm{f(M_{1})}=11.0 \pm 2.1 \,\rm{M_{\odot}}$ \citep{Daniel_2015}, we can analytically estimate the mass of the companion star $M_{2}$. We establish a range that falls between $ 0.0006 \pm 0.0001  \, \rm{M_{\odot}}$ and $0.32 \pm 0.06 \, \rm{M_{\odot}}$ (note that \citealt{Daniel_2015} reported a lower limit of $M_{2}\, > 0.4 \,M_{\odot}$).
In other words, the inclination estimated for this system together with the limits on the mass ratio are in line with the literature, and the mass range for the companion star $M_{2}$ characterises it as a low-mass star that fills its Roche lobe.


\subsubsection{Inner and outer radius of the emitting region}
We found that the \hb{} line in J1357 is emitted from a region bounded between $(0.16-0.91)\, R_{T}$, while the \ha{} emission line observed in the J1305 spectrum describes two different emitting regions for outburst and quiescence. During the quiescent phase, the emission line is located between $(0.1-1.3) \,R_{T}$, while during the outburst, the emitting region extends to radii between  $(0.83-13) \,R_{T}$ and $(1.1-19)\, R_{T}$.

 Table \ref{tab_diskline} shows that the outer radius of J1357 
is smaller than the expected tidal radius, indicating that either the accretion disc does not extend to the tidal radius or that the \hb{} emission line does not originate in the outermost region of the accretion disc.
For J1305, the best-fit value for the outer radius obtained during quiescence is consistent with the expected geometry of the system (although the best-fit range includes values that exceed the tidal radius). On the other hand, the outer radius obtained for the outburst spectra is entirely out of scale.
A value of about 13-19 $ \, R_{T}$ is higher by up to an order of magnitude than the orbital separation. This deviation is not due to a miscalculation or to a limitation of the application of \textsc{diskline} under certain circumstances. 
\cite{Miceli2023_submitted} reported a Doppler-shift velocity of the peaks during the outburst of approximately $400 \, \rm{km \, s^{-1}}$. Considering a Keplerian disc, we can apply the third Keplerian law to calculate the corresponding radius, and even in this case, the derived radius is four times the tidal radius.  
This result could be due to the effects during the outburst on the line profile.
It is possible that the velocity distribution of the emitting region is not Keplerian. 
In standard discs \citep{Shakura_Sunyaev}, the in-flowing material reaches a stable equilibrium in which the radial component of the velocity is negligible compared to the azimuthal component. However, if there is a surge in viscosity within that region, the radial velocity may cease to be negligible, and the way in which matter accretes may deviate from the way in the Keplerian regime.

An alternative explanation might be linked to the existence of a circumbinary disc around the binary system. Due to the high-accretion rate, outflows of matter can occur, and the ejected material may subsequently remain around the system, causing the formation of a viscous toroidal disc around the binary system \citep{Chen_2019ApJ...876L..11C}.
Although these discs are typically quite cold, the intense emission produced by the central source ($\rm{L_{x} \sim 10^{37} \, erg\, s^{-1}}$ for J1305 during outburst, see \citealt{Miller_2014ApJ...788...53M}) might irradiate these regions, thus providing the energy to trigger the atomic transitions that give rise to the \ha{} emission line. This would explain an emitting region greater than the orbital separation itself. 
However, these are only hypothetical answers to a question that requires further analysis to be understood.

\subsection{Temperatures and hydrogen ionisation}
Although our results seem to delineate an emitting region that is consistent with the geometry of the disc, at least during quiescence periods, it is crucial to consider the temperatures at these radii. It is indeed true that the emission lines \hb{} and \ha{} are associated with the energy transitions of neutral hydrogen. This means that when the temperatures are too high, the amount of neutral hydrogen will be negligible.

\cite{Shakura_Sunyaev} described the structure of the accretion disc around a BH under the assumption that the disc is optically thick and geometrically thin. The authors divided the disc into three ideal regions, depending on the predominant type of pressure and cross-section, and derived a system of equations for each of them through which the physics of the accretion disc can be completely described. 
We analysed the outer part of the disc, and the equation that describes the variation in the central temperature of the disc with the radius therefore is
\begin{equation}\label{temperature}
    T=8.6 \cdot 10^7 \alpha^{-1 / 5} \dot{m}^{3 / 10} m^{-1 / 5} r^{-3 / 4}\left(1-r^{-1 / 2}\right)^{3 / 10}, 
\end{equation}
where  $m,\dot{m}, and r$ are non-dimensional parameters that contain the dependence on the BH mass, the accretion rate, and the radius, respectively (see \cite{Shakura_Sunyaev} for the explicit dependence), while $\rm{\alpha}$ is the accretion disc viscosity parameter, which usually lies in the range 0.1- 0.4 \citep{2007_King_alpha}. In this case, we assumed  $\rm{\alpha} =0.1$. As the density does not vary significantly across the vertical height ($z$) at these large radii, it can reasonably be assumed that the surface temperature of the disc is approximately equal to the central temperature \citep{Shakura_Sunyaev}.

Considering the values of $\rm{R_{in}}$ and $\rm{R_{out}}$ from our analysis, we estimated the temperature ranges for the \hb{} and \ha{} emitting regions during the quiescent phases of J1357 and J1305, as we have already discussed the reliability of the measurements during the outburst. We obtained temperature ranges of $2081 \pm 255$ K to $353 \pm 40$ K for J1305 and $2203 \pm 317$ K to $580 \pm 81$ K for J1357. This is consistent with a higher quiescent luminosity of $\rm{L_{X} \sim 1.6 \times 10^{32}, erg, s^{-1}}$ for J1305 (\citealt{2021_mata-sanchez}) compared to J1357, for which $\rm{L_{X} \sim 1.3 \times 10^{31}, erg, s^{-1}}$ (\citealt{Armas_2014}).


The degree of ionisation for the H-atom, that is, the fraction of ionised hydrogen with respect to neutral hydrogen at the estimated temperatures, can be computed using the Saha equation \citep{Payne_1924Natur.113..783P},
\begin{equation}\label{saha}
    \log \rm{\frac{H^{+}}{H}}=\log \frac{u^{+}}{u}+\log 2+\frac{5}{2} \log T- 5040\, \frac{\chi_{\mathrm{ion}}}{T}-\log P_e-0.48.
\end{equation} 
Here, $\rm{P_e = n_e kT}$ is the electron pressure, where $\rm{n_{e}}$ is the electron density, which is typically assumed to be equal to $10^{18}\,\rm{cm^{-3}}$ \citep{Sincell_1998}. $\chi_{\mathrm{ion}}$ is the ionization energy (13.6 eV for the H atom), and $\rm{u}$ is the partition function of the atom ($\rm{u(H)}= 2$ and $\rm{u(H^{+})}=1$ (see \citealt{Vitense_1992isa..book.....B}).
Assuming the temperatures we inferred above, we obtain a range of $\rm{H^{+}/H}$ between $10^{-25}$ and $10^{-114}$ for J1357, and between $10^{-45}$ and $10^{-186}$ for J1305. This implies that the amount of ionised hydrogen with respect to neutral hydrogen is negligible, and therefore, the radii estimated for the emitting regions of the \ha{} and \hb{} lines for the two systems appear to be physically plausible.

\section{Conclusion}
We presented a new approach for analysing the emission lines found in the optical spectra of binary systems that involves the use of the \textsc{diskline} model in order to obtain constraints on the system geometry.

We analysed two observations in quiescence of the X-ray binary BH candidates J1357 and J1305  and an observation collected during an outburst of the latter source.
The best-fit parameters allowed us to provide a reasonable description of the geometry of these systems.
The \hb{} and \ha{} emission lines in the quiescent spectra of J1357 and J1305 are emitted by a ring in the disc between $(0.16 - 0.91) \, R_{T}$ and $(0.1-1.3)\, R_{T}$, respectively. 
The analysis of the outburst spectra yielded an emission region that is not aligned with the expected system geometry, suggesting a range for the outer radius from the tidal radius to values exceeding the orbital separation of the binary system.
We can put forward some hypotheses to account for this behaviour, including 
a non-Keplerian flow in the outer disc or the existence of a circumbinary disc.

Our analysis reveals that the inclination angles of the two systems closely match the expected values. This confirms their high-inclination nature.
The proposed method requires further investigation because the physics of emission lines from accretion discs is far more complex, and the \textsc{diskline} model provides a simplified description of the emission profile modified by Doppler effects in a Keplerian flow, without any consideration of self-absorption effects in the optically thick atmosphere of accretion discs.
However, the application of the \textsc{diskline} model to the disc emission lines in the optical band can be a powerful tool for providing a  reliable geometrical description of these sources, allowing us to give precise estimates of the inclination angle, and also to gather information on the expected temperature and ionisation level of the emitting region.
Further analyses similar to the analysis we presented here will be useful to provide additional evidence regarding the reliability of this method and to fully explore its capabilities.
For example, applying the method to an eclipsing source would represent an optimal approach to test its validity. It would also be interesting to ascertain whether different emission lines within the same spectrum yield different insights. In theory, access to a diverse array of emission lines would allow us to delineate the emission from distinct rings of the disc.

It is clear, therefore, that obtaining credible constraints on the parameters of the accretion disc is a challenging task. The application of this method might provide a compatibility check which together with the results that have been obtained with other methods might shed some light on the disc geometry.

\begin{acknowledgements}
The authors acknowledge the financial contribution 
PRIN-INAF 2019 with the project "Probing the geometry of accretion: from theory to observations" (PI: Belloni). 

\end{acknowledgements}

\bibliography{46909corr}
\bibliographystyle{aa}

\begin{appendix}

\section{Additional table}
\begin{table}[h]
 \begin{adjustbox}{angle=270, minipage=\textwidth}
 \caption{\label{tab_diskline} Best-fit values for the parameters of the two models described in the text that include the \textsc{diskline} component. Uncertainties are at the 90\% c.l.}
\begin{threeparttable}
  \resizebox{1.2\textwidth}{!}{\begin{minipage}{\textwidth}
\renewcommand{\arraystretch}{1.4} 

\begin{tabular}{ccccccccccc}
\hline
  & \textsc{spectra}                & \multicolumn{6}{c}{\textsc{diskline}}                                                & \multicolumn{3}{c}{\textsc{gaussian}}                        \\ 

\hline
&   & $\lambda$ (\AA) & Index & R$_{in}$(10$^{3}$ $R_{g}$) & R$_{out}$(10$^{4}$ $R_{g}$) & Inc (degrees)& N & $\lambda$ (\AA) & $\sigma$ (\AA) &N\\ \hline 

{\sc J1357} & src1-Q1 & $4861.274^{+0.005}_{-0.002}$
 & $-2.34 ^{+0.06}_{-0.07}$
& $9.6 \pm 0.2$ & $5.7 \pm 0.2$ & $>80$ & $104 \pm 4$& $4861.07 \pm 0.03$
 & $4.322 \pm 0.005$ & $-8 \pm 2$   \\
 
   & src1-Q2 & $4715.592^{+0.085}_{-0.123}$ & *&* &*&*&  $79 \pm 3$ & $4717.49 \pm 0.12$
 &$3.79 \pm 0.04$  & $-2^{+2}_{-1}$
 \\ 
    & src1-Q3 & $4854.175^{+0.005}_{-0.035}$& *& * &*&*&$ 81  \pm 3$ & $4854.84 \pm 0.05$
 &$4.944 \pm 0.017$  &$-3.4^{+1.1}_{-2.1}$
 \\
    & src1-Q4 & $4859.217^{+0.005}_{-0.004}$ &* &* &*&*& $76^{+2}_{-3}$  & $4859.43 \pm 0.03$
 & $4.983 \pm 0.006$ &$-5 \pm 1$
 \\
    & src1-Q5 & $4860.284^{+0.012}_{-0.027}$ &* & * &*&*& $75 \pm 3$  & $4858.89 \pm 0.04$
& $4.952 \pm 0.011$ &$-4.1  \pm 1.5$
\\
    & src1-Q6 & $4857.827^{+0.009}_{-0.006}$&* &* &*&*& $77^{+3}_{-4}$ &  $4855.56 \pm 0.05$
 & $2.68 \pm 0.01$& $-2.7 \pm 1.3$
\\
    & src1-Q7 & $4857.446^{+0.003}_{-0.006}$
 & *&* &*&*& $85 \pm 4$  & $4857.58 \pm 0.03$
 & $4.968 \pm 0.008$ &$-6 \pm 2$
\\
     & src1-Q8&  $4858.817^{+0.002}_{-0.008}$
 &*&* &*&*& $103 \pm 4$  & $4858.76 \pm 0.03$
 &$4.980 \pm 0.006$ & $-8 \pm 2$  \\
\hline 
  & $\chi^2/dof$ & 2781.81/2787 \\
  \hline

{\sc J1305} & src2-Q1 &$6561.8715^{+0.0009}_{-0.0011}$ & $-1.62^{+0.02}_{-0.03}$ & $19.4^{+0.7}_{-0.9}$ & $20.1 \pm +0.4$ & $72.6^{+1.4}_{-1.3}$ & $38.4^{+1.2}_{-2.0}$ &  $6562.9830^{+0.0006}_{-0.0017}$ & $2.7922 \pm 0.0008$ & $-5.8^{+0.4}_{-0.8}$ \\ 

{\sc quiet} & src2-Q2 & $6563.2262 \pm 0.0012 $ & * & * & * & * & $34 \pm 1$ &$6562.9483^{+0.0015}_{-0.0007}$  & $1.9820 \pm 0.0008$
& $-3.4^{+0.4}_{-0.5} $
\\ 
    & src2-Q3 & $6564.1992^{+0.0011}_{-0.0006}$ & * & * & * & * & $35.5^{+1.2}_{-0.7}$ &$6562.9136 \pm 0.0009$
 &$2.0555 \pm 0.0007$
 &$-3.3^{+0.3}_{-0.5} $ \\
    & src2-Q4 &$6563.6779^{+0.0009}_{-0.0035}$  & * & * & * & * & $37.5^{+0.8}_{-0.9}$ &$6563.9907^{+0.0008}_{-0.0010}$
 &$1.4805 \pm 0.0007$ &$-2.2^{+0.4}_{-0.3} $ \\
     & src2-Q5& $6563.9907 \pm 0.0007$  & * & * & * & * & $40\pm 1$ &$6564.2687^{+0.0012}_{-0.0008}$
 &$2.6952 \pm 0.0007$ & $-4.3 \pm 0.4 $\\

    & src2-Q6 & $6563.1568^{+0.0006}_{-0.0005}$ & * & * & * & * & $39.6^{+1.1}_{-0.8}$ &$6561.5589^{+0.0005}_{-0.0009}$
&$1.9395 \pm 0.0006$ &$-3.3 \pm 0.3 $
\\
    & src2-Q9 &$6562.5662^{+0.0011}_{-0.0009}$
  & * & * & * & * & $39.8^{+0.7}_{-1.1}$ &$6562.531^{+0.001}_{-0.002}$
 &$2.9834 \pm 0.0012$ &$-5.6^{+0.5}_{-0.9} $
\\
    & src2-Q10& $6561.4548 \pm 0.0009 $& * & * & * & * & $38 \pm 1 $ & $6562.5314^{+0.0014}_{-0.0008}$ &$2.1840 \pm 0.0008$ &$-3.4 \pm 0.4 $
 \\

    & src2-Q15&  $6565.7288^{+0.0008}_{-0.0006}$  & * & * & * & * & $43.0_{-1.8}^{+0.9}$ &$6565.659^{+0.053}_{-0.003}$
 &$1.6578 \pm 0.0047$ & $-1.8^{+0.3}_{-0.8} $\\
     & src2-Q16 &$6563.0873^{+0.0011}_{-0.0008}$  & * & * & * & * &$38 \pm 1$    &$6561.4548^{+0.0009}_{-0.0012}$
 &$3.6891 \pm 0.0009$ &$-8.3^{+0.6}_{-0.8} $
\\

 \hline 
  &   $\chi^2/dof$ & 2620.0/2620 \\
  \hline

{\sc J1305} & src2-O2 & $6565.0682^{+0.0009}_{-0.0009}$ & $-2.4 \pm 0.1$& $120^{+9}_{-7}$ & $196^{+32}_{-27}$ & $70 \pm 4$& $1.6 \pm 0.1$ & -& -& - \\ 
{\sc day 1} & src2-O4 & $6565.1377^{+0.0008}_{-0.0019}$& * &* &* & * & $ 1.35 \pm 0.09$& -& -& -
\\ 
            & src2-O5 & $6566.042^{+0.002}_{-0.001}$ &* &* &* & * &$1.21 \pm 0.09$& -& -& -\\
             & src2-O6 & $6565.173^{+0.002}_{-0.001}$ & * &* &* & * &$1.20  \pm 0.09$& -& -& -\\
        \hline 
         & $\chi^2/dof$ & 139.4/153 \\
         \hline
{\sc day 2} & src2-O7 & $6565.103 \pm 0.001$ & $-2.7 \pm 0.2$ & $166^{+10}_{-8}$ & $273^{+90}_{-60}$ & $71 \pm 4$ & $1.16^{+0.11}_{-0.08}$ &$6587.953^{+0.002}_{-0.002}$ & $4.901 \pm 0.002$ & $-0.24^{+0.09}_{-0.06}$
\\ 
            & src2-O8 & $6566.077^{+0.002}_{-0.001}$& *&  *& *& *&$0.9^{+0.11}_{-0.08}$ & *& *&$-0.16^{+0.05}_{-0.10}$
 \\
             & src2-O9 & $6563.886^{+0.094}_{-0.002}$& *&  *& *& *&$0.9^{+0.11}_{-0.09}$ & *& *&$-0.17^{+0.07}_{-0.08}$
\\
             & src2-O10 &$6563.574^{+0.001}_{-0.002}$& *&  *& *& *&$1.11 ^{+0.07}_{-0.15}$ & *& *&$-0.33^{+0.11}_{-0.05}$
\\
             & src2-O11 & $6564.686^{+0.002}_{-0.001}$& *&  *& *& *&$1.11^{+0.11}_{-0.09}$ & *& *&$-0.23 ^{+0.09}_{-0.06}$
 \\ 
             & src2-O12 & $6564.929^{+0.001}_{-0.001}$& *&  *& *& *&$1.17 \pm 0.09$ & *& *&$-0.20^{+0.09}_{-0.06}$
 \\
\hline

 & $\chi^2/dof$ & 149.0/222 \\
 \hline
\end{tabular}
\begin{tablenotes}
    \item[*]Kept linked to the first data set during the fit.
\end{tablenotes} \end{minipage}}
\end{threeparttable}
\end{adjustbox}
\end{table}
\end{appendix}
\end{document}